%Paper: 9204006
%From: "HILL, BRIAN" <bhill@physics.ucla.edu>
%Date: 16 Apr 92 11:59:00 PDT

%The figures for this paper were not produced electronically, but copies
%will be mailed upon request to oscarh@physics.mcgill.ca or bhill@uclaph.
%
%changed \sendingaddress to \address and eliminated \end from the definition
%of \endofletter for compatibility with lab.tex and addbkfmt.tex
%%%%%%%%%%%%%%%%%%%%%%% Letter and Preprint Macro %%%%%%%%%%%%%%%%%%%%%%%%%
\hsize=6.0truein\vsize=8.5truein\voffset=0.25truein
\hoffset=0.1875truein%would be 0.25 except our laser printer is off by 1/16in
\tolerance=1000\hyphenpenalty=500
\def\monthintext{\ifcase\month\or January\or February\or
   March\or April\or May\or June\or July\or August\or
   September\or October\or November\or December\fi}

\def\monthdayandyear{\monthintext\space\number\day, \number\year}

%%%%%%%%%%%%%%%%%%%%%%%  Twelve point text font  %%%%%%%%%%%%%%%%%%%%%%%%%%
\font\tenrm=cmr10 scaled \magstep1   \font\tenbf=cmbx10 scaled \magstep1
\font\sevenrm=cmr7 scaled \magstep1  
\font\fiverm=cmr5 scaled \magstep1   
\font\teni=cmmi10 scaled \magstep1   \font\tensy=cmsy10 scaled \magstep1
\font\seveni=cmmi7 scaled \magstep1  \font\sevensy=cmsy7 scaled \magstep1
\font\fivei=cmmi5 scaled \magstep1   \font\fivesy=cmsy5 scaled \magstep1

\font\tentt=cmtt10 scaled \magstep1
\font\tenit=cmti10 scaled \magstep1
\font\tensl=cmsl10 scaled \magstep1
\def\twelvepoint{\def\rm{\fam0\tenrm}
   \textfont0=\tenrm \scriptfont0=\sevenrm \scriptscriptfont0=\fiverm
   \textfont1=\teni  \scriptfont1=\seveni  \scriptscriptfont1=\fivei
   \textfont2=\tensy \scriptfont2=\sevensy \scriptscriptfont2=\fivesy
   \textfont\itfam=\tenit \def\it{\fam\itfam\tenit}
   \textfont\ttfam=\tentt \def\tt{\fam\ttfam\tentt}
   \textfont\bffam=\tenbf \def\bf{\fam\bffam\tenbf}
   \textfont\slfam=\tensl \def\sl{\fam\slfam\tensl} \rm
   %Essentially I changed all dimensions to 1.2 times as large as in plain tex
   \hfuzz=1pt\vfuzz=1pt%much more than plain tex's value
   \setbox\strutbox=\hbox{\vrule height 10.2pt depth 4.2pt width 0pt}
   \parindent=24pt\parskip=1.2pt plus 1.2pt
   \topskip=12pt\maxdepth=4.8pt\jot=3.6pt
   \normalbaselineskip=13.5pt\normallineskip=1.2pt
   \normallineskiplimit=0pt\normalbaselines
   \abovedisplayskip=13pt plus 3.6pt minus 5.8pt
   \belowdisplayskip=13pt plus 3.6pt minus 5.8pt
   \abovedisplayshortskip=-1.4pt plus 3.6pt
   \belowdisplayshortskip=13pt plus 3.6pt minus 3.6pt
   %ignore plain tex's value for belowdisplayshortskip looked terrible
   \topskip=12pt \splittopskip=12pt
   \scriptspace=0.6pt\nulldelimiterspace=1.44pt\delimitershortfall=6pt
   \thinmuskip=3.6mu\medmuskip=3.6mu plus 1.2mu minus 1.2mu
   \thickmuskip=4mu plus 2mu minus 1mu%reduced these plain tex values
   \smallskipamount=3.6pt plus 1.2pt minus 1.2pt
   \medskipamount=7.2pt plus 2.4pt minus 2.4pt
   \bigskipamount=14.4pt plus 4.8pt minus 4.8pt}
\twelvepoint

%%%%%%%%%%%%%%%%%%%%%%%  Definitions for Letters  %%%%%%%%%%%%%%%%%%%%%%%%
\def\letter{\parindent=0pt\parskip=\medskipamount\def\endmode{}
   \def\longindent{\parindent=3.25truein\obeylines\parskip=0pt}
   \def\letterhead{\null\vfil\begingroup
      \parindent=3.25truein\obeylines\parskip=0pt
      \def\endmode{\medskip\medskip\endgroup}}
   \def\date{\endmode\begingroup\parindent=3.25truein\obeylines\parskip=0pt
      \def\endmode{\medskip\medskip\endgroup}\monthdayandyear}
   \def\address{\endmode\begingroup
      \parindent=0pt\obeylines\parskip=0pt
      \def\endmode{\medskip\medskip\endgroup}}
   \def\salutation{\endmode\begingroup
      \parindent=0pt\obeylines\parskip=0pt\def\endmode{\medskip\endgroup}}
   \def\body{\endmode\begingroup\parskip=\medskipamount
      \def\endmode{\medskip\medskip\endgroup}}
   \def\closing{\endmode\begingroup\longindent
      \def\endmode{\endgroup}}
   \def\signed{\endmode\begingroup\longindent\vskip0.8truein
      \def\endmode{\endgroup}}
   \def\endofletter{\endmode \ifnum\pageno=1 \nopagenumbers\fi
      \vfil\vfil\eject}}

%%%%%%%%%%%%%%%%%%% Definitions for Preprint Title Page %%%%%%%%%%%%%%%%%%%%
\def\preprint#1{ %use with \draft or \date to finish title page
   \def\draft{\finishtitlepage{PRELIMINARY DRAFT:
\monthdayandyear}\writelabels%
      \headline={\sevenrm PRELIMINARY DRAFT: \monthdayandyear\hfil}}
   \def\date##1{\finishtitlepage{##1}}
   \font\titlerm=cmr10 scaled \magstep3
   \font\titlerms=cmr10 scaled \magstep1 %\font\titlermss=cmr8
   \font\titlei=cmmi10 scaled \magstep3  %math italic for title
   \font\titleis=cmmi10 scaled \magstep1 %\font\titleiss=cmmi8
   \font\titlesy=cmsy10 scaled \magstep3 	%math symbols for title
   \font\titlesys=cmsy10 scaled \magstep1  %\font\titlesyss=cmsy8
   \font\titleit=cmti10 scaled \magstep3	%text italic for title
   \skewchar\titlei='177 \skewchar\titleis='177 %\skewchar\titleiss='177
   \skewchar\titlesy='60 \skewchar\titlesys='60 %\skewchar\titlesyss='60
   \def\titlefont{\def\rm{\fam0\titlerm}% switch to title font
      \textfont0=\titlerm \scriptfont0=\titlerms %\scriptscriptfont0=\titlermss
      \textfont1=\titlei  \scriptfont1=\titleis  %\scriptscriptfont1=\titleiss
      \textfont2=\titlesy \scriptfont2=\titlesys %\scriptscriptfont2=\titlesyss
      \textfont\itfam=\titleit \def\it{\fam\itfam\titleit} \rm}
   \def\title##1{\vskip 0.7in plus 0.4in\centerline{\titlefont ##1}}
   \def\authorline##1{\vskip 0.7in plus 0.4in\centerline{\bf ##1}
      \vskip 0.12in plus 0.02in}
   \def\author##1##2##3{\vskip 0.9in plus 0.4in
      \centerline{{\bf ##1}\myfoot{##2}{##3}}\vskip 0.12in plus 0.02in}
   \def\addressline##1{\centerline{##1}}
   \def\abstract{\vskip 0.5in plus 0.35in\centerline{\bf Abstract}\smallskip}
   \def\finishtitlepage##1{\vskip 0.8in plus 0.4in
      \leftline{##1}\supereject\endgroup}
   \baselineskip=18.8pt plus 0.2pt minus 0.2pt \pageno=0
   \begingroup\nopagenumbers\parindent=0pt\baselineskip=13.5pt\rightline{#1}}

%%%%% Numbered Sections, Equations, Footnotes, References and Figures %%%%%%
\def\nolabels{\def\eqnlabel##1{}\def\eqlabel##1{}\def\figlabel##1{}%
   \def\reflabel##1{}}
\def\writelabels{\def\eqnlabel##1{%
   {\escapechar=` \hfill\rlap{\hskip.11in\string##1}}}%
   \def\eqlabel##1{{\escapechar=` \rlap{\hskip.11in\string##1}}}%
   \def\figlabel##1{\noexpand\llap{\string\string\string##1\hskip.66in}}%
   \def\reflabel##1{\noexpand\llap{\string\string\string##1\hskip.37in}}}
\nolabels
%  tagged section numbers
\global\newcount\secno \global\secno=0
\global\newcount\meqno \global\meqno=1
\def\newsec#1{\global\advance\secno by1
   \xdef\secsym{\the\secno.}
   \global\meqno=1\bigbreak\medskip
   \noindent{\bf\the\secno. #1}\par\nobreak\smallskip\nobreak\noindent}
\xdef\secsym{}
\def\appendix#1#2{\global\meqno=1\xdef\secsym{\hbox{#1.}}\bigbreak\medskip
\noindent{\bf Appendix #1. #2}\par\nobreak\smallskip\nobreak\noindent}
\def\acknowledgements{\bigbreak\medskip\centerline{\bf
   Acknowledgements}\par\nobreak\smallskip\nobreak\noindent}
%  equations
\def\eqnn#1{\xdef #1{(\secsym\the\meqno)}%
	\global\advance\meqno by1\eqnlabel#1}
\def\eqna#1{\xdef #1##1{\hbox{$(\secsym\the\meqno##1)$}}%
	\global\advance\meqno by1\eqnlabel{#1$\{\}$}}
\def\eqn#1#2{\xdef #1{(\secsym\the\meqno)}\global\advance\meqno by1%
	$$#2\eqno#1\eqlabel#1$$}
%  footnotes
\def\myfoot#1#2{{\baselineskip=13.5pt plus 0.3pt\footnote{#1}{#2}}}
%sequentially numbered footnotes
\global\newcount\ftno \global\ftno=1
\def\foot#1{{\baselineskip=13.5pt plus 0.3pt\footnote{$^{\the\ftno}$}{#1}}%
	\global\advance\ftno by1}
%  references
\global\newcount\refno \global\refno=1
\newwrite\rfile
\def\ref{[\the\refno]\nref}
\def\nref#1{\xdef#1{[\the\refno]}\ifnum\refno=1\immediate
   \openout\rfile=\jobname.aux\fi\global\advance\refno by1\chardef\wfile=\rfile
   \immediate\write\rfile{\noexpand\item{#1\ }\reflabel{#1}\pctsign}\findarg}
\def\findarg#1#{\begingroup\obeylines\newlinechar=`\^^M\passarg}
   {\obeylines\gdef\passarg#1{\writeline\relax #1^^M\hbox{}^^M}%
   \gdef\writeline#1^^M{\expandafter\toks0\expandafter{\striprelax #1}%
   \edef\next{\the\toks0}\ifx\next\null\let\next=\endgroup\else\ifx\next\empty%

\else\immediate\write\wfile{\the\toks0}\fi\let\next=\writeline\fi\next\relax}}
   {\catcode`\%=12\xdef\pctsign{%}}
\def\striprelax#1{}
\def\semi{;\hfil\break}
\def\addref#1{\immediate\write\rfile{\noexpand\item{}#1}} %now unnecessary
\def\listrefs{\vfill\eject\immediate\closeout\rfile
   \centerline{{\bf References}}\medskip{\frenchspacing%
   \catcode`\@=11\escapechar=` %
   \input \jobname.aux\vfill\eject}\nonfrenchspacing}
\def\startrefs#1{\immediate\openout\rfile=refs.tmp\refno=#1}
%  figures
\global\newcount\figno \global\figno=1
\newwrite\ffile
\def\fig{\the\figno\nfig}
\def\nfig#1{\xdef#1{\the\figno}\ifnum\figno=1\immediate
   \openout\ffile=\jobname.fig\fi\global\advance\figno by1\chardef\wfile=\ffile
   \immediate\write\ffile{\medskip\noexpand\item{Fig.\ #1:\ }%
   \figlabel{#1}\pctsign}\findarg}
\def\listfigs{\vfill\eject\immediate\closeout\ffile{\parindent48pt
   \baselineskip16.8pt\centerline{{\bf Figure Captions}}\medskip
   \escapechar=` \input \jobname.fig\vfill\eject}}

%%%%%%%%%%%%%%%%%%%%%%%%%%%%%%%%%%%%%%%%%%%%%%%%%%%%%%%%%%%%%%%%%%%%%%%%%%%
\def\noblackbox{\overfullrule=0pt}
\def\inv{^{\raise.18ex\hbox{${\scriptscriptstyle -}$}\kern-.06em 1}}
\def\dup{^{\vphantom{1}}}
\def\Dsl{\,\raise.18ex\hbox{/}\mkern-16.2mu D} %this one can be subscripted
\def\dsl{\raise.18ex\hbox{/}\kern-.68em\partial}
\def\slash#1{\raise.18ex\hbox{/}\kern-.68em #1}
\def\boxeqn#1{\vcenter{\vbox{\hrule\hbox{\vrule\kern3.6pt\vbox{\kern3.6pt
   \hbox{${\displaystyle #1}$}\kern3.6pt}\kern3.6pt\vrule}\hrule}}}
\def\mbox#1#2{\vcenter{\hrule \hbox{\vrule height#2.4in
   \kern#1.2in \vrule} \hrule}}  %e.g. \mbox{.1}{.1}
\def\bar{\overline}\def\psibar{\bar\psi}
\def\e#1{{\rm e}^{\textstyle#1}}
\def\del{\partial}
\def\curly#1{{\hbox{{$\cal #1$}}}}
\def\curlyL{\hbox{{$\cal L$}}}
\def\vev#1{\langle #1 \rangle}
\def\lform{\hbox{$\sqcup$}\llap{\hbox{$\sqcap$}}}
\def\darr#1{\raise1.8ex\hbox{$\leftrightarrow$}\mkern-19.8mu #1}
\def\half{{\textstyle{1\over2}}} %puts a small half in a displayed eqn
\def\roughly#1{\ \lower1.5ex\hbox{$\sim$}\mkern-22.8mu #1\,}
\def\MSbar{$\bar{{\rm MS}}$}
\hyphenation{di-men-sion di-men-sion-al di-men-sion-al-ly}
\def\ket#1{\vert #1 \rangle}\def\bra#1{\langle #1 \vert}
\def\Half{{1\over2}}
\def\d#1#2{d\mskip 1.5mu^{#1}\mkern-2mu{#2}\,}
\def\mev{\mathop{\rm Me\kern-0.1em V}\nolimits}
\def\gev{\mathop{\rm Ge\kern-0.1em V}\nolimits}
\def\alphaS{{\alpha_S}}\def\alphapi{{\alpha_S\over4\pi}}
\def\npb#1,#2,#3 {Nucl.\ Phys.\ {\bf B#1} (19#2) #3}
\def\prd#1,#2,#3 {Phys.\ Rev.\ D {\bf #1} (19#2) #3}
\def\plb#1,#2,#3 {\ifnum#1>170 Phys.\ Lett.\ B {\bf #1} (19#2) #3\else
 Phys.\ Lett.\ {\bf #1B} (19#2) #3\fi}
\def\sjnp#1,#2,#3 {Sov.\ J.\ Nucl.\ Phys.\ {\bf #1} (19#2) #3}
\def\yadfiz#1,#2,#3 {Yad.\ Fiz.\ {\bf #1} (19#2) #3}
\baselineskip=18pt plus 0.2pt minus 0.2pt %took away 1pt
\def\curlyO{\curly O}
\def\alphabeta{\alpha\thinspace\thinspace\thinspace\thinspace\beta}
\def\bdagger{b^\dagger}
\def\fsubB{$f_B$}

\noblackbox
\preprint{hep-lat/9204006}
\rightline{UCLA/91/TEP/51}\rightline{MCGILL/91--39}
\vskip 0.5in plus 0.1in
\title{\titlefont Point-Split Lattice Operators for B Decays}
\vskip 0.85in plus 0.4in
\centerline{\bf Oscar F. Hern\'andez}
\vskip 0.12in plus 0.02in
\addressline{Department of Physics }
\addressline{McGill University }
\addressline{Ernest Rutherford Physics Building }
\addressline{Montr\'eal, Qu\'e., Canada H3A 2T8 }
\vskip 0.2in plus 0.02in
\centerline{\bf Brian R. Hill}
\vskip 0.12in plus 0.02in
\addressline{Department of Physics}
\addressline{University of California}
\addressline{Los Angeles, CA~~90024}
\abstract
The matrix element which determines the $B$ meson decay constant can
be measured on the lattice using an effective field theory for heavy
quarks.  Various discretizations of the heavy-light bilinears
which appear in this and other $B$ decay matrix elements are
possible.  The heavy-light
bilinear currently used for the determination of the $B$ meson decay
constant on the lattice suffers a substantial one-loop
renormalization.  In this paper, we compute the one-loop
renormalizations of the discretizations in which the heavy and light
fields in the bilinear are separated by one lattice spacing, and discuss
their application.
\date{12/91, to appear in Physics Letters B.}
%%%%%%%%%%%%%%%%%%%%%%%%%%%%%%%%%%%%%%%%%%%%%%%%%%%%%%%%%%%
\newsec{Introduction}%
Several weak matrix elements involving heavy mesons can be studied on
the lattice~\ref\Etalk{E. Eichten, in {\sl Field
Theory on the Lattice,} edited by A.~Billoire {\it et al.}, Nucl.
Phys. B (Proc. Suppl.) {\bf 4} (1988) 170.}\ref\LandTtalk{G. P.
Lepage and B. A. Thacker, {\it ibid}, p. 199.}.
The relationship of the lattice
operators to the operators coming from the continuum electroweak
theory must be calculated in order to make use of the lattice results.  While
these short-distance strong interaction corrections are in
principle perturbatively calculable, in practice, the one-loop
corrections are sometimes so large as to be of questionable reliability.
The uncertainty in whether one should use the lattice or continuum value
of the strong coupling in the one-loop computation
and the unknown magnitude of higher loop corrections are both
measures of this uncertainty.  For the operator currently in use to
determine the $B$ meson decay constant on the lattice, these factors
yield an uncertainty of as much as twenty per cent.

There is actually considerable freedom in the choice of the lattice
operators.  They are generally
chosen for convenience in calculating the matrix element on the
lattice and in performing the perturbative renormalization.
However, the choice may also be made so as to minimize systematic
effects, such as effects which vanish as a power of the lattice spacing,
or the effect of perturbative corrections.  For the fermion
bilinear determining the decay constants of the
light pseudoscalar mesons, the use of a
point-split operator substantially changes the perturbative
correction~\ref\MandZ{G. Martinelli and Y.-C. Zhang, \plb 125,83,77.}.
Similarly, it has been suggested that a bilinear suitable
for the determination of the $B$ meson decay constant that is distance-one
could have a smaller perturbative correction than the distance-zero
operator commonly in use~\ref\pbm{P. B. Mackenzie, private communication.}.
In this paper, we compute the renormalization of the distance-one
bilinears which are suitable for determining $f_B$ and other $B$ decay matrix
elements and  which respect the symmetries of the lattice heavy quark action.

The calculation uses techniques developed and applied to calculate
renormalizations of other lattice operators involving heavy quarks~%
\ref\fb{E. Eichten and B. Hill, \plb240,90,193.}%
--\nref\hnh{O. F. Hern\'andez and B. R. Hill, \plb237,90,95.}%
\ref\ff{J. M. Flynn, O. F. Hern\'andez and B. R. Hill, Phys. Rev. D {\bf
43} (1991) 3709.}.
In the following section, we review the lattice regularization of the
heavy quark
effective field theory and the result for the renormalization of
the operator currently being used to determine $f_B$.
In order to concentrate on the new results and the possible applications of
the new operators, we will be brief in reviewing
the background for the computation and will refer the reader to the appropriate
references for the technical details.  Section three contains the
calculation of the one-loop renormalization of an operator with
splitting in the spatial directions.  In section four, operators
with splitting in the time direction are considered.
In the conclusion, we discuss the possible applications of these results.
%%%%%%%%%%%%%%%%%%%%%%%%%%%%%%%%%%%%%%%%%%%%%%%%%%%%%%%%%%%%%%%%%%%%
\newsec{Zero-Distance Bilinears}%
An expansion which analytically removes the dependence on the $b$ quark
mass~\hbox{\ref\eikonal{J. M. Cornwall and G. Tiktopoulos, Phys. Rev. D
{\bf 15} (1977) 2937.}%
\ref\expansion{E. Eichten and F. Feinberg, Phys. Rev. D {\bf 23} (1981) 2724.}}
can be used to determine matrix elements of $B$
mesons on the lattice~\Etalk\LandTtalk.
The relationships between operators in
the full theory and their counterparts in the effective theory built around the
zeroth order term in this
expansion \LandTtalk\ref\CaswellAndLepage{W. E. Caswell and G.
P. Lepage, \plb167,86,437 .}\nref\PandWII{H. D. Politzer and M. B.
Wise, \plb208,88,504 .}\nref\eft{E. Eichten and B. Hill,
\plb234,90,511 .}\nref\Grin{B. Grinstein, Nucl. Phys. {\bf B339} (1990) 253.}%
\nref\Georgi{H.~Georgi, \plb240,90,447 .}--\ref\ggd{M. J. Dugan, M.
Golden, and B. Grinstein, HUTP/91--A045.}\ are perturbatively calculable.

Many discretizations of the heavy quark effective theory action,
\eqn\Leff{S=\int {d\,}^4x\,\bdagger(i\partial_0+gA_0) b,}
with the same naive continuum limit are possible.
The choice of action,
\def\zerohat{\hat 0}
\eqn\S{S=i a^3 \sum_n\left[b^\dagger(n)\left(
b(n)-U_0(n{-}\zerohat)^\dagger b(n{-}\zerohat)
\right)\right],}
reproduces the heavy quark propagator used in numerical simulations.
In the preceding two equations, $b$ is a two-component column vector
which annihilates heavy quarks, $A_0=A_0^aT_a$, is the continuum
$SU(3)$ gauge field, $a$ is the lattice spacing, and
$U_0=\exp(igaA_0)$ is the lattice gauge field living on time links.

The most general heavy-light bilinear in the full theory is
\eqn\J{J(x)=\bar b(x) \Gamma q(x).}
Here $\Gamma$ is any Dirac matrix, and $q$ is the light quark field.
In a Dirac basis, parametrize $\Gamma$ by two-by-two blocks:
\eqn\matrix{\Gamma=\left(
{\alphabeta\atop\gamma\thinspace\thinspace\thinspace\thinspace\delta}
\right)\!.}
In the heavy quark effective theory, the operator~\J\ takes the form
\eqn\cont{\bdagger(x)(\alphabeta)q(x),}
where $(\alphabeta)$ is a two-by-four matrix.
An obvious choice of discretization of this operator which has the
right continuum limit is the zero distance bilinear,
\eqn\obvious{b^\dagger(n)(\alphabeta)q(n).}
The light quark is treated as a Wilson fermion.  Staggered and naive
fermion results can be obtained by setting the Wilson mass parameter
to zero~\hnh.

In order to parametrize the relationship between the continuum
operator and the lattice operator, we introduce $c$-numbers $G$
and~$H$, defined by $G\Gamma=\gamma_0\Gamma\gamma_0$ and
$H\Gamma=\gamma_\mu\Gamma\gamma_\mu$.  We also need
$H'$, the derivative with respect to $d$ of $H$ in $d$ dimensions.
This derivative depends on
the extension of the gamma matrix algebra in the dimensionally regularized
theory.  With these definitions,
the ratio of the continuum effective theory operator to the
operator in the full theory is~\eft\ref\blp{Ph. Boucaud, C. L. Lin and
O. P\`ene, Phys. Rev. D {\bf 40} (1989) 1529\semi
Ph. Boucaud, C. L. Lin, and O. P\`ene, Phys. Rev D {\bf 41} (1990) 3541(E).}
\eqn\effectivetofull{1-{g^2\over12\pi^2}
\left(C_1\ln{m^2\over \mu^2}+C_2\right)\!,}
where $C_1=5/2{-}H^2\!/4$ and $C_2={-}4{+}3H^2\!/4{-}HH'{-}GH/2$.
For the case of $\Gamma=\gamma_0\gamma_5$,
the bilinear used to determine \fsubB,
$G=-1$ and $H=2$.
If we use the extension of the gamma matrix algebra that $\gamma_5$
anticommutes with all the $\gamma_\mu$, $1\le\mu\le d$, then $H'=1$.
So in this case, we have $C_1=3/2$ and $C_2=-2$.

We also need the ratio of zero distance bilinear to the operator in
the continuum effective theory.  It is~\fb\blp
\eqn\zerotoeffective{1+{g^2\over12\pi^2}\left[
d+{1\over2}e+{1\over2}f-{5\over4}
%\left(D+{1\over2}E+{1\over2}F\right)
\right]\!.}
The dependence on $\mu a$ has been eliminated by setting $\mu=1/a$.
The constant
\eqn\donedtwo{
d=d_1+d_2G,}
arises from the vertex correction graphs on
the lattice. When we consider the spatially split operator in the next section,
we will find that its effect will be to decrease the value of $d_1$ and $d_2$.
The constant $e=24.48$ arises from heavy quark wave function
renormalization on the lattice.  This constant should be reduced
to 4.53 if
one extracts $f_B$ from lattice results by fitting to $A\e{-Bn_0a}$
rather than $A\e{-B(n_0{+}1)a}$~\fb.
The constant $f$ arises from light
quark wave function renormalization on the lattice~\ref\MandZI{G.
Martinelli and Y.-C. Zhang, \plb123,83,433.}.
The constants $d_1,$ $d_2$ and $f$ which depend on the Wilson mass
coefficient $r$, have been calculated and tabulated previously~\fb\blp,
and analytical expressions can be found in reference~\ff.
For completeness, we tabulate these constants in Table~1 for several
values of $r$. Errors for any of the numerically evaluated constants are at
most $\curlyO(1)$ in the last decimal place.

\topinsert
\vbox{
\def\tablerule{\noalign{\hrule}}
$$\vbox{\offinterlineskip\tabskip=0em
	\halign to 4in{\vrule #\tabskip=1em plus5em&
		\strut\hfil$#$\hfil&\vrule #&
      \hfil$#$\hfil&\vrule #&
      \hfil$#$\hfil&\vrule #&
      \hfil$#$\hfil&\tabskip=0em\vrule #\cr
		\tablerule
%		height 2pt&\omit&&&&&&&\cr
		&r&&d_1&&d_2&&f&\cr
%		height 2pt&\omit&&&&&&&\cr
		\tablerule
		&1.00&&5.46&&-7.22&&13.35&\cr
		\tablerule
		&0.75&&5.76&&-7.23&&11.96&\cr
		\tablerule
		&0.50&&6.30&&-7.00&&10.22&\cr
		\tablerule
		&0.25&&7.37&&-5.72&&\phantom{1}8.07&\cr
		\tablerule
		&0.00&&8.79&&\phantom{-}0.00&&\phantom{1}6.54&\cr
		\tablerule}}
$$
\centerline{Table 1.}}
\endinsert

We illustrate the use of these results for~$r=1.00$ and~%
$a^{-1}=2\gev$. With~$\Gamma=\gamma_0\gamma_5$, the case
of interest in calculating~$f_B$,
the constant~$G$, which appears in equation~\donedtwo\ is~$-1,$
as noted below equation~\effectivetofull.
Taking~$\mu=2\gev$ and the~$b$~quark mass~$m=5\gev$, with~$\alphaS=0.25$,
the effective to
full theory continuum ratio~\effectivetofull\ is~$0.98$. %more precisely 0.980
The bare lattice value of~$g^2$ at~$a^{-1}=2\gev$ is approximately~$1.0$.
For the perturbative matching however we will take~$g^2=1.8$ which is roughly
the continuum value at the scale~$\pi/a$~\ref\unconv{G. P.
Lepage and P. B. Mackenzie, in {\sl Lattice '90,} ed. by U. M. Heller
{\it et al,} Nucl. Phys. B (Proc. Suppl.) {\bf 20} (1991) 173.}.
With this choice
the ratio of the lattice to continuum operators is~1.31.  The value of~%
$f_B$ extracted from numerical simulations from the first of the two
fitting procedures mentioned above should therefore be divided by~1.28,
which is the product of these two ratios.  This reduction
factor would be~1.43 if the unreduced value of $e$ were used.
%%%%%%%%%%%%%%%%%%%%%%%%%%%%%%%%%%%%%%%%%%%%%%%%%%%%%%%%%%%%%%%%%%
\newsec{Spatially-Split Bilinears}%
A variety of other operators with the same continuum limit as the operator in
equation \obvious\ can be constructed.  In this section, we will consider the
distance-one bilinear with splitting in the spatial directions given by
\def\ihat{\hat \imath}
\eqn\spac{{1\over6}\sum_i\left[
\bdagger(n{+}\ihat)U_i(n)^\dagger(\alphabeta)q(n)+
\bdagger(n{-}\ihat)U_i(n{-}\ihat)(\alphabeta)q(n)
\right].}
The index $i$ runs only over the three spatial directions.
The sum of six terms has been chosen to
respect the remnants of the $O(3)$
rotational group present in the lattice heavy quark effective theory.

It is simplest to compute the relationship of this operator to the
continuum operator~\J\ by computing its relationship to the
zero-distance bilinear~\obvious.  Thus we will give analytical and
numerical values
for the change in the constants $d_1$ and $d_2$, quoted in the
previous section.  There are only four
graphs to compute. There is the vertex correction graph which gave
rise to $d_1$ and
$d_2$ in the case of the zero distance bilinear, Figure~\fig\vertexfig{The
vertex correction graph which is present for all heavy-light bilinears.}.
It is unaffected in appearance, but its value must be
recomputed due to factors that arise in the spatially split case.
There are also three new vertex correction graphs,
which are depicted in Figure~\fig\newgraphs{Three new vertex
correction graphs (present for the distance-one bilinears).}.
Graph~(a) in the figure vanishes
in Feynman gauge because
the gauge link at the vertex is spatial and the heavy quark field
only interacts with the time component of the gauge field.

In order to quote the analytic expression for these graphs we define:
\def\summu{\sum\nolimits_\mu}
\eqn\Deltas{\eqalign{\Delta_1&=\summu\sin^2{l_\mu\over2},\cr
\Delta_4&=\summu\sin^2{l_\mu}, \cr
\Delta_2&=\Delta_4+4r^2\Delta_1^2 .}}
The sums on $\mu$ run from 1 to 4.
$\Delta_1^{(3)}$ and $\Delta_2^{(3)}$ are defined to be identical to
$\Delta_1$ and $\Delta_2$ respectively except with $l_4$ set to zero.

The Feynman integral arising from the difference of the vertex
correction in the case of the spatially-split operator~\spac\
and the zero-distance operator~\obvious\ is
\def\hpd{{-i(e^{il_0}-1)+i\epsilon}}
\def\lpn{\sum_\sigma\gamma_\sigma\sin l_\sigma-2ir\Delta_1}
\eqn\one{-{4\over3}g^2(\alphabeta)
	\int {d^4l\over(2\pi)^4}
{(\lpn)\,\e{il_0/2}\left(\gamma_0\cos l_0/2+ir\sin l_0/2\right)\Delta_1^{(3)}
\over 6\Delta_1 \Delta_2(\hpd) }.}
This is identical to the expression for the vertex correction for the
Feynman integrand for the operator~\obvious, except for an additional factor
of $-2\Delta_1^{(3)}/3$.  This factor
necessarily vanishes for small $l$ since the two operators have the
same naive continuum limit.

The second graph in Figure~\newgraphs\ gives
\eqn\two{
{4\over3}g^2(\alphabeta)
\int {d^4l\over(2\pi)^4}{
\left(\lpn\right)
{1\over3}\sum_i\sin{l_i/2}
\left(\gamma_i\cos{l_i/2}+ir\sin{l_i/2}\right)
\over 4\Delta_1\Delta_2}.}
The integral for the graph in Figure~\newgraphs (c), the tadpole graph, is
frequently encountered.  It is
\eqn\frog{-{4\over3}\,g^2(\alphabeta)
	\int {d^4l\over(2\pi)^4}{1\over8\Delta_1}.}

We evaluate these integrals using the techniques for
isolating non-covariant poles of
reference~\fb.  The sum of all graphs gives a change in $d_1$ of:
\eqn\deltadone{
\Delta d_1= -{1\over\pi^2}	\int {d^4l}
{ 4\Delta_1-\Delta_1^2 + 2\Delta_4 +12r^2\Delta_1^2
  \over 24 \Delta_1 \Delta_2 }
}
The contribution to $d_2$ comes from the part of the graphs that is
proportional to~$\gamma_0$. Only the graph of Figure~\vertexfig\
(whose Feynman integral appears in equation~\one) has a contribution
of this type.  The contribution is odd in the Wilson parameter.
We find,
\eqn\deltadtwo{\Delta d_2={r\over3}{1\over\pi}\int d^3l
{\Delta_1^{(3)}\over \Delta_2^{(3)}}.}
These expressions for $\Delta d_1$ and $\Delta d_2$
have been evaluated using VEGAS~\ref\VEGAS{G. P.
Lepage, J. Comp. Phys. {\bf 27} (1978) 192.}, and are tabulated in
Table~2 for various values of the Wilson mass coefficient $r$.
Errors are at most $\curlyO(1)$ in the last decimal place.

\topinsert
\def\tablerule{\noalign{\hrule}}
%\vskip\abovedisplayskip
\vbox{
$$\vbox{\offinterlineskip\tabskip=0em
	\halign to 4in{\vrule #\tabskip=1em plus5em&
		\strut\hfil$#$\hfil&\vrule #&
      \hfil$#$\hfil&\vrule #&
      \hfil$#$\hfil&\tabskip=0em\vrule #\cr
		\tablerule
%		height 2pt&\omit&&&&&&&\cr
		&r&&\Delta d_1&&\Delta d_2&\cr
%		height 2pt&\omit&&&&&&&\cr
		\tablerule
		&1.00&&-12.70&&3.89&\cr
		\tablerule
		&0.75&&-12.98&&4.44&\cr
		\tablerule
		&0.50&&-13.49&&4.98&\cr
		\tablerule
		&0.25&&-14.62&&4.81&\cr
		\tablerule
		&0.00&&-16.30&&0.00&\cr
		\tablerule}}
$$
\centerline{Table 2. Change in $d_1$ and $d_2$ versus Wilson Mass Parameter.}}
\endinsert
As in the previous section, we illustrate the use of these results for
$r=1.00$, $a^{-1}=2\gev$, $g^2=1.8$, and $\Gamma=\gamma_0\gamma_5$.
The effective to full theory ratio of $0.98$ remains unchanged.
With $\Delta d_1=-12.70$ and
$\Delta d_2=3.89$, $d$ is reduced to $-3.91.$  Then
the ratio of the lattice to continuum operators is 1.06.  %more precisely 1.057
The value of $f_B$ extracted from numerical simulations from
the spatially-split bilinear fitting to $A\e{-Bn_0a}$
should therefore be divided by
1.04 (the product of these two ratios).  %more precisely 1.036
As at the end of the previous section,
this product would be larger, 1.18, if one fits to $A\e{-B(n_0{+}1)a}.$
%%%%%%%%%%%%%%%%%%%%%%%%%%%%%%%%%%%%%%%%%%%%%%%%%%%%%%%%%%%%%%%%%%
\newsec{Temporally-Split Bilinears}%
In this section, we will consider the two
distance-one bilinears with splitting
either forward or backward in the time direction:
\eqn\timeplus{b^\dagger(n{+}\zerohat)U_0(n)^\dagger(\alphabeta)q(n),}
and
\eqn\timeminus{b^\dagger(n{-}\zerohat)U_0(n{-}\zerohat)(\alphabeta)q(n).}
Except for variations in the way the gauge links could be inserted between the
bilinears, this exhausts the
possible distance-one bilinears which respect the remnants of the
$O(3)$ rotational group that is present in the lattice heavy quark
effective theory.  We will see that
these operators do not lead to any new correlation functions on
the lattice.  They do however, highlight the problem of
extracting decay constants in the presence of the linearly divergent
heavy quark self-energy.

As in the previous section we compute the relationship of these operators to
the continuum operator~\J\ by computing its relationship to the
distance zero bilinear~\obvious.
The consideration of the new Feynman diagrams is somewhat similar to the
discussion for the spatially split bilinears discussed in section
three, and we will proceed directly to the contributions of the
various vertex correction graphs.  The graph which gave rise to $d_1$ and
$d_2$ (Figure~\vertexfig) must again be recomputed.  However, we find that its
contribution is cancelled by the contribution of graph~(b)
in Figure~\newgraphs.  This is true for both of the temporally split
bilinears.
The contribution of the remaining two graphs, (a) and (c)
in Figure~\newgraphs, is
\eqn\deltad{\Delta d_1=\pm{1\over\pi}\int d^3l
{1\over4\Delta_1^{(3)}}.}
The upper sign is for the operator in equation \timeplus~and
the lower sign is for the operator in
equation~\timeminus.  Numerically, this integral has the value $19.95.$
For either operator we find $\Delta d_2=0.$

When one evaluates correlation functions of either of these operators on the
lattice, a paradox appears: there is absolutely no change in the combination of
gauge links and light quark propagators that is used in the temporally-split
case as compared to the zero distance case.  For the operator \timeminus, the
final gauge link is associated with the operator instead of the propagator, but
the product is unchanged. A similar thing occurs for the operator \timeplus,
except that there one additional link and its inverse appear in the product and
cancel. The lack of a change in the correlator is disturbing since we have
found a change in the relationship of the operator to its continuum
counterpart. For simplicity, in what follows we will just discuss the latter of
the two temporally-split operators~\timeminus.

The resolution of this paradox brings us back to a subtlety already
explored in the
renormalization of the distance-zero operator \fb.  Because the
propagator used in numerical simulations of heavy quarks corresponds to an
unrenormalized action, and because the heavy quark self-energy is
linearly divergent, it matters whether one fits the correlation
function measured in numerical simulations to $A\e{-Bn_0a}$ or to
$A\e{-B(n_0{+}1)a}$. As explained in reference~\fb,
for the distance zero operator, the latter
exponential is the natural one to fit to.
One can repeat the argument for the
temporally split distance one bilinear~\timeminus.  In this case we
find
that the {\it former} exponential is the natural one to fit to.

It was also noted in reference~\fb\ that if one  uses the
former exponential with the zero-distance bilinear~\obvious,
it can be compensated for by reducing $e$ by an amount
taken from the linearly divergent part of the heavy quark self energy.%
\myfoot{$^\dagger$}{The correlator being discussed in reference~\fb, had
both the zero-distance
bilinear~\obvious\ and its conjugate, and the error of fitting to the
former exponential was split between the two operators.  If it had
just been associated with one of the two operators, the error in
fitting to the former exponential could have been compensated for by a
reduction in $d_1$.}  Perturbatively, the
amount of the reduction was $19.95$--exactly
the amount by which we have found $d_1$ to be reduced for the
operator~\timeminus.
A similar argument can be given for the operator \timeplus.

The resolution of the apparent paradox relied on a perturbative estimate for
the linearly divergent part of the heavy quark self energy.
The use of the operators \obvious, \timeplus\ and \timeminus\ fitted to
the appropriate exponentials will only
give the same results to the extent that a perturbative estimate
of the linear divergence of the heavy quark self energy is accurate.
%%%%%%%%%%%%%%%%%%%%%%%%%%%%%%%%%%%%%%%%%%%%%%%%%%%%%%%%%%%%%%%%%%
\newsec{Conclusions}%
We have renormalized three distance-one bilinears chosen to respect the
symmetries of the lattice heavy quark action.  Our results can be
used whether the light quark is treated as a Wilson, staggered or
naive fermion.
These operators can be applied in a variety of ways.  Most
fundamentally, they provide a basic consistency check for this
approach to determining $B$ decay matrix elements.  A lattice
calculation to check for agreement of the results should be undertaken.

After basic consistency is established, one can imagine
three ways of applying our results for the spatially-split bilinear.
This operator had a significantly smaller correction
than the correction to the bilinear that is
currently used for the determination of $f_B$ (whether the reduced or
unreduced value of the heavy quark wave function renormalization is used).
Hence that part of the uncertainty which is due to the choice
of the value of the strong coupling is correspondingly reduced.
Optimistically, one could then assume that this is the more accurately
renormalized operator and simply use the resulting values.  Further, one
could assume that the results for both operators are accurate
and fix the value of $g^2$ by choosing it to eliminate any
disagreement between them.  A third more conservative application would be to
take the difference in the values for $f_B$, with a range of
reasonable values for $g^2$, as an estimate of the systematic error
resulting from perturbative matching and choice of discretization.

The application of the various temporally-split operators has been
shown to be related to
studying the dependence on extracting $f_B$ from lattice simulations
using various exponential fitting functions.
The two temporally-split operators highlight the fact that the perturbative
estimate for the heavy quark self energy must be accurate or else
different operators with the same naive continuum limit will give
results that disagree.  The values of the
parameters in these fitting functions and their statistical
correlation must be reported to study
this dependence.  This will give us another estimate of some of
the systematic errors in this approach to computing $f_B$.
%%%%%%%%%%%%%%%%%%%%%%%%%%%%%%%%%%%%%%%%%%%%%%%%%%%%%%%%%%%%%%%
\acknowledgements
We thank Estia Eichten for discussions.
OFH was supported in part by the National Science and Engineering Research
Council of Canada, and les Fonds FCAR du Qu\'ebec.
\listrefs
%\listfigs
\bye